\documentclass[aps,pra,twocolumn,showpacs,superscriptaddress]{revtex4}
\usepackage{graphicx}
\usepackage{epstopdf}
\usepackage{amsmath}
\usepackage{amssymb}
\usepackage{amsfonts}
\DeclareMathOperator{\Tr}{Tr}
\usepackage{dcolumn}

\begin{document}

\title{Small 
two-component
Fermi gases in a cubic box with periodic boundary conditions}

\author{X. Y. Yin}
\affiliation{Department of Physics and Astronomy, Washington State University, Pullman, Washington 99164 USA}
\author{D. Blume}
\affiliation{Department of Physics and Astronomy, Washington State University, Pullman, Washington 99164 USA}
\affiliation{ITAMP, Harvard-Smithsonian Center for Astrophysics,
Cambridge, Massachusetts 02138, USA}
\date{\today}

\begin{abstract}
The properties of two-component Fermi gases become 
universal if the interspecies 
$s$-wave 
scattering length 
$a_s$ and the average interparticle spacing 
are much larger than the range of the underlying two-body potential. 
Using an explicitly correlated Gaussian basis set expansion approach, 
we determine the eigen energies of two-component Fermi gases 
in a cubic box with periodic boundary conditions 
as functions of the interspecies $s$-wave scattering length 
and the effective range of the two-body potential. 
The universal properties of systems 
consisting of up to four particles 
are determined by extrapolating the finite-range energies 
to the zero-range limit. 
We determine the eigen energies of states 
with vanishing and finite momentum. 
In
the 
weakly-attractive 
BCS regime, 
we analyze the energy  
spectra 
and 
degeneracies
using  
first-order 
degenerate perturbation theory.
Excellent agreement between the perturbative energy shifts and
the numerically determined energies is obtained.
For the infinitely large scattering length case, 
we compare our 
results---where available---with 
those presented in the literature. 
\end{abstract}


\maketitle

\section{Introduction}
Two-component Fermi gases 
with interspecies
contact interactions have emerged as a paradigm 
of strongly-correlated 
systems~\cite{giorginireview,blochreview,kett08,doertereview,gezerlis}.
A detailed understanding of the equation of state of two-component
Fermi gases as functions of the strength of the contact interaction and
the temperature is, e.g., of importance to 
nuclear and astrophysics.
Dilute ultracold atomic $^{6}$Li and $^{40}$K gases provide nearly ideal 
table-top realizations of 
this paradigm system.
Indeed, much of our current understanding of strongly-correlated
Fermi systems throughout the BCS-BEC crossover and
at unitarity comes from a host of
experimental cold atom studies.
These experimental studies are complemented by 
theoretical studies.

When the $s$-wave scattering length, which can be tuned
through the application of an external magnetic 
field in the vicinity of a Fano-Feshbach 
resonance~\cite{chinreview},
becomes large, the system does not possess a small parameter and
non-perturbative approaches are needed.
In this regime, the equation of state of two-component Fermi gases
has been determined by Monte
Carlo as well as other non-perturbative 
methods~\cite{giorginireview,carlson,giorgini,bulgac,burovski,juillet,AFQMC,kaplan}. 
While the fixed-node diffusion Monte
Carlo method yields variational upper bounds, other
Monte Carlo methods are expected to
provide, within the statistical uncertainty, 
essentially exact 
results~\cite{carlson,giorgini,bulgac,burovski,AFQMC}. 
In assessing the accuracy of the various theoretical approaches,
exact diagonalization schemes of small model systems
play a crucial role~\cite{deanlee}.
The
explicitly correlated Gaussian
basis set expansion approach has been used 
extensively to treat two-component Fermi gases 
under harmonic 
confinement~\cite{sorenson,stec07c,blum07,blume1,blume2,blum11,debraj12}.
The present work extends the standard
explicitly correlated Gaussian approach~\cite{suzuki,RMPreview}
to study few-body systems in a cubic box 
with periodic boundary conditions. 
The method introduced in this paper 
is directly applicable to other periodic systems
such as atoms loaded into optical lattices.
In addition to serving as a benchmark, our
study of strongly-correlated 
few-body systems in a cubic box with periodic
boundary conditions aids in developing a physical understanding of
the corresponding 
many-body systems. The results are also
relevant to the analysis of on-going lattice QCD 
simulations~\cite{castin,BeanSavage,beane}.

This work considers equal-mass Fermi 
gases consisting of $N_1$ spin-up fermions 
and $N_2$ spin-down fermions
in a cubic box of length $L$
with periodic boundary conditions.
We consider the regime where the unlike particles do not
interact and where the interspecies interactions are
characterized by a short-range potential with $s$-wave scattering length
$a_s$ and effective range $r_{\text{eff}}$.
The key points of this paper are: 
{\em{(i)}} We introduce explicitly correlated 
Gaussian basis functions and show that the resulting basis,
constructed using the stochastic variational approach~\cite{svm},
provides an accurate description of few-body states
with vanishing and non-vanishing momentum.
Compact analytic expressions for the most important matrix elements are 
reported.
{\em{(ii)}} We analyze the energy spectra
and degeneracies of the  
$(N_1, N_2)=(1,1)$, $(2,1)$, $(2,2)$ and $(3,1)$ systems 
in the weakly-attractive BCS regime (i.e., for $|a_s|/L \ll 1$ and $a_s <0$)
using first-order degenerate perturbation theory.
{\em{(iii)}} 
Tables~\ref{table_unitarity11}-\ref{table_unitarity22and31} 
present accurate results
for the ground state energies of the
$(N_1, N_2)= (1,1)$, $(2,1)$, $(2,2)$ and $(3,1)$ systems, 
and the excited states of the
$(1,1)$ and $(2,1)$ systems at unitarity. 
Our extrapolated zero-range energy of the $(2,2)$
system is in excellent agreement with earlier benchmark 
results~\cite{deanlee}.
{\em{(iv)}} We present energy spectra throughout the BEC-BCS crossover.

The remainder of this paper is organized as follows.  
Section~\ref{sec_theory} discusses the theoretical framework. 
Specifically,
Sec.~\ref{sec_systemham}
introduces the system Hamiltonian,
Sec.~\ref{sec_pt} 
discusses 
the degenerate perturbation theory treatment 
of the
weakly-attractive BCS regime, and 
Sec.~\ref{sec_ecg} 
introduces the explicitly correlated Gaussian basis set
expansion approach.
Section~\ref{sec_results} presents 
our
results
for the $(1,1)$, $(2,1)$, $(2,2)$ and $(3,1)$ systems. 
Lastly, 
Sec.~\ref{sec_sumary}
concludes. 
Details of the 
explicitly correlated Gaussian basis functions
for 
systems
with periodic boundary conditions are relegated to 
the
Appendix.

\section{Theoretical framework}
\label{sec_theory}
\subsection{System Hamiltonian}
\label{sec_systemham}
We study equal-mass two-component Fermi gases 
consisting of
$N_1$ spin-up and $N_2$ 
spin-down atoms ($N=N_1+N_2$) 
in a cubic box of length $L$
with periodic boundary conditions. 
Besides the box, the particles feel no external forces.
The system Hamiltonian $H$ reads
\begin{equation}\label{general Hamiltonian}
H=H_0+V_{\text{int}}, 
\end{equation} 
where $H_0$,
\begin{equation} \label{NI Hamiltonian}
H_0=\sum_{a=1}^{N}-\frac{\hbar^2}{2m}\nabla_a^2,
\end{equation}
is the non-interacting Hamiltonian and $V_{\text{int}}$,
\begin{equation} \label{interaction}
V_{\text{int}}=\sum_{a=1}^{N_1} \sum_{b=N_1+1}^{N}
V_{\text{tb}}(\mathbf{x}_{ab}),
\end{equation}
is the pairwise additive interaction potential.
In Eq.~(\ref{NI Hamiltonian}),
$m$ denotes the atom
mass and
$\nabla_a^2$ the Laplacian of the $a$th atom
with position vector $\mathbf{x}_a$.
The two-body interaction potential $V_{\rm{tb}}$ depends on the
interparticle distance vector $\mathbf{x}_{ab}$,
where
$\mathbf{x}_{ab}=\mathbf{x}_{a}-\mathbf{x}_{b}$.

We consider two different short-range model potentials 
$V_{\text{tb}}(\mathbf{x}_{ab})$.
Our perturbative treatment
(see Sec.~\ref{sec_pt}) 
employs the bare 
or non-regularized
Fermi pseudopotential $V_\text{F}$~\cite{ferm34}, 
\begin{equation} \label{fermi potential}
V_{\text{F}}(\mathbf{x}_{ab})=\frac{4 \pi \hbar^2 a_s}{m} 
\delta^{(3)} (\mathbf{x}_{ab}),
\end{equation}
where $a_s$ is 
the
two-body 
free-space
$s$-wave scattering length. 
Our 
explicitly correlated Gaussian basis set expansion
approach 
(see Secs.~\ref{sec_ecg} and \ref{sec_results}), in contrast,
employs
a finite-range
Gaussian potential $V_\text{g}$ 
with range $r_0$ and depth $U_0$,
\begin{equation} \label{gauss potential}
V_{\text{g}}(\mathbf{x}_{ab})=
U_{0} \exp \left( -
\frac{\mathbf{x}_{ab}^2}{2r_{0}^2}\right)
.
\end{equation}
For a fixed $r_0$, $U_0$ 
($U_0<0$)
is adjusted 
to generate potentials with different $a_s$.
Throughout, we restrict ourselves to two-body potentials
that support zero and one two-body $s$-wave bound states
in free space for $a_s$ negative and positive, respectively.
 
\subsection{Perturbative treatment}
\label{sec_pt} 
In the
weakly-attractive 
regime, i.e., for $|a_s|/L \ll 1$ ($a_s<0$), we treat the two-component
Fermi gas 
in a cubic box 
with periodic boundary conditions perturbatively.
Specifically, the potential $V_{\text{int}}$ with $V_{\text{tb}}=V_{\text{F}}$,   
see Eqs.~(\ref{interaction}) and (\ref{fermi potential}), 
is treated as 
a perturbation to 
the non-interacting Hamiltonian 
$H_0$,
Eq.~(\ref{NI Hamiltonian}).
Using standard 
first-order
degenerate time-independent perturbation theory,
we determine the leading-order 
energy shifts
of  
the non-interacting energy levels and corresponding degeneracies.
Moreover, we 
construct properly anti-symmetrized eigen states 
that simultaneously diagonalize $H_0$ and the total
momentum operator.

The unsymmetrized eigen states $\Phi^{(0)}$
of the unperturbed Hamiltonian $H_0$,
Eq.~(\ref{NI Hamiltonian}),
with periodic boundary conditions
are most conveniently written in terms of plane 
wave states,
\begin{eqnarray}
\Phi
^{(0)}
_{\mathbf{k}_{1}, 
\cdots
, \mathbf{k}_{N}} 
( 
\mathbf{x}_{1}, 
\cdots
, \mathbf{x}_{N})
=
\frac{1}{L^{3N/2}}
\prod_{a=1}^{N}
\exp ( \imath \mathbf{k}_a  
\cdot
\mathbf{x}_a ),
\end{eqnarray}
where the wave vectors $\mathbf{k}_a$ 
satisfy the condition
\begin{equation}
\mathbf{k}_a=\frac{2 \pi}{L} \mathbf{n}_a
\end{equation}
with $\mathbf{n}_a=(n_a^{(1)},n_a^{(2)},n_a^{(3)})$
and $n_a^{(j)}=0,1,\cdots$.
The corresponding unperturbed eigen energies $E_n^{(0)}$
read
\begin{eqnarray}
\label{eq_eni}
E_{n}^{(0)}=n E_{\text{box}},
\end{eqnarray}
where
\begin{equation}
E_{\text{box}}=\frac{2\pi^{2} \hbar^{2} }{m L^{2}}
\end{equation}
and
\begin{eqnarray}
\label{eq_nqn}
n = \sum_{a=1}^N \mathbf{n}_a^2.
\end{eqnarray}
As can be seen from Eqs.~(\ref{eq_eni}) and (\ref{eq_nqn}),
the energies $E_n^{(0)}$ are, except for the lowest state with
$n=0$, degenerate.

We
obtain the energy 
shifts 
$\Delta E_{n,q}$ of the energy level $E_n^{(0)}$
by diagonalizing the matrix
\begin{equation}
\label{eq_perturbationmatrix}
\langle
\Phi^{(0)}_{\mathbf{k}_{1}, 
\cdots
, \mathbf{k}_{N}}
|V_{\text{int}}|\Phi^{(0)}
_{\mathbf{k}_{1}', 
\cdots
, 
 \mathbf{k}_{N}'}
\rangle
,
\end{equation}
which is constructed using
the unperturbed states $\Phi^{(0)}_{\mathbf{k}_1,\cdots,\mathbf{k}_N}$ 
and
$\Phi^{(0)}_{\mathbf{k}_1',\cdots,\mathbf{k}_N'}$ with energy $E^{(0)}_n$. 
As a result of the interaction,
each degenerate non-interacting energy 
$E_n^{(0)}$ 
is split into $Q$
sublevels with distinct energy shift
$\Delta E_{n,q}$ ($q=1,\cdots,Q$) and $|\mathbf{K}|$ (see below).
The non-interacting states that diagonalize the
perturbation matrix,
Eq.~(\ref{eq_perturbationmatrix}), 
are also eigenstates of the total momentum
operator $\mathbf{P}$,
\begin{eqnarray}\label{momentum operator}
\mathbf{P}=-\imath \hbar \sum_{a=1}^{N} 
\nabla_a,
\end{eqnarray}
with eigenvalue $\hbar \mathbf{K}$,
where 
$\mathbf{K}=\sum_{a=1}^N \mathbf{k}_a$.

Up to this point, no symmetry constraints have been imposed.
To construct states with proper fermionic exchange symmetry,
we form all possible linear combinations of states 
that satisfy the anti-symmetry 
requirement under the interchange of 
pairs of identical fermions
for each manifold labeled by 
$\Delta E_{n,q} $ and $\hbar |\mathbf{K}|$.

Table~\ref{table_deg}
summarizes 
the energy shifts $\Delta E_{n,q}$,
the magnitude of the total momentum $\hbar |\mathbf{K}|$,
and the corresponding degeneracies for the lowest few
states of the 
$(1,1)$, $(2,1)$, $(2,2)$ and $(3,1)$ 
Fermi
systems.
\begin{table}
\caption{Perturbative treatment of 
$(1,1)$, $(2,1)$, $(2,2)$ and $(3,1)$ systems.
Column two shows the non-interacting energy $E_n^{(0)}$.
Columns three and four report the first-order energy shift
$\Delta E_{n,q}$ and the 
magnitude of the total wave vector $|\mathbf{K}|$.
The degeneracy of each state is shown in
column five.}
\label{table_deg}
\centering
\begin{tabular}{c|cccr}
\hline
\hline
 $(N_1, N_2)$  &  $E_n^{(0)}/E_{\text{box}}$  & $\Delta E_{n,q}/ (\frac{4\pi\hbar^2a_s}{mL^3})$  
 &  $|\mathbf{K}|/(\frac{2 \pi}{L}) $ & Deg. \\
 \hline
 (1, 1)  & 0 & 1 & 0 & 1 \\
 \cline{2-5}
           & 1 & 2 & 1 & 6 \\
           &    & 0 & 1 & 6 \\
 \cline{2-5}
           & 2 & 6 & 0 & 1 \\
           &    & 4 & $\sqrt{2}$ & 12 \\
           &    & 1 & 2 & 6 \\
           &    & 0 & 0 & 5 \\
           &    & 0 & $\sqrt{2}$ & 36 \\     
 \cline{1-5}             
 (2, 1)  & 1 & 2 & 1 & 6 \\ 
 \cline{2-5}
           & 2 & 7 & 0 & 1 \\
           &    & 4 & $\sqrt{2}$ & 12 \\
           &    & 3 & 0 & 3 \\
           &    & 3 & $\sqrt{2}$ & 12 \\
           &    & 2 & 2 & 6 \\
           &    & 1 & 0 & 2 \\
           &    & 1 & $\sqrt{2}$ & 12 \\
           &    & 0 & 0 & 3 \\
           &    & 0 & $\sqrt{2}$ & 12 \\
 \cline{1-5}
 (2, 2)  & 2 & 9 & 0 & 1 \\
           &    & 5 & $\sqrt{2}$ & 12 \\
           &    & 4 & 2 & 6 \\
           &    & 3 & 0 & 5 \\
           &    & 3 & $\sqrt{2}$ & 12 \\
 \cline{1-5}
 (3, 1)  & 2 & 3 & 0 & 3 \\
           &    & 3 & $\sqrt{2}$ & 12 \\         
\hline
\hline
\end{tabular}
\end{table} 
For the $(1,1)$ system, the lowest non-interacting 
state is one-fold degenerate and is characterized by a perturbation shift
of $\Delta E_{n,q}=4 \pi \hbar^2 a_s/(mL^3)$ and 
magnitude of total momentum of $\hbar |\mathbf{K}|=0$.
The first excited and second excited non-interacting states, in
contrast, are 12-fold and 60-fold degenerate, respectively.
The degeneracy of 12 arises since 
plane wave states with $\mathbf{K}/(2\pi/L)=(\pm 1,0,0)$,
$(0,\pm 1,0)$ and $(0,0,\pm 1)$ are degenerate.
Moreover, the state with $\mathbf{K}/(2\pi/L)=(1,0,0)$, e.g.,
can be constructed by putting either the first or the second particle
into the first excited state, yielding a total degeneracy of 12.
The interactions split
the first excited state into two levels with
degeneracy six each. One level is shifted down by the 
attractive interactions,
while the other is unshifted, reflecting
the fact that the wave function vanishes when the two
particles sit on top of each other.
The second excited state, which has a degeneracy of 60 in the absence of
interactions, is split into five levels with distinct 
$\Delta E_{n,q}$ and $|\mathbf{K}|$ ``labels''.

For $N=3$ and $4$, the counting of the degeneracies 
is  more involved than for the $(1,1)$ system,
since the (unperturbed)
non-interacting wave functions have to be anti-symmetric under
the exchange of identical fermions.
The fact that the non-interacting ground state of the 
$(2,1)$, $(2,2)$ and $(3,1)$ systems has a finite 
energy, and not a vanishing energy
as in the $(1,1)$ case, is a direct consequence of the 
fermionic anti-symmetry requirement.
Another interesting aspect of the results summarized
in Table~\ref{table_deg} is that the ground state of
the $(2,1)$ system has, in the weakly-attractive regime,
a finite momentum while the ground state of the
$(2,2)$  system has a vanishing momentum.
Interestingly, the first-order perturbation theory shift
of the lowest two levels of the $(3,1)$ system,
which have vanishing and finite momentum, respectively,
is identical. The momentum of the
true ground state in the BCS regime can thus not be
determined within first-order perturbation theory but
requires the determination of higher-order corrections or the usage
of a 
non-perturbative technique.

The perturbative treatment breaks down when $|a_s|/L$ is not
small compared to $1$. To treat systems with arbitrary 
$s$-wave scattering length $a_s$,
we resort to a numerical approach, the explicitly correlated Gaussian approach.

\subsection{Explicitly correlated Gaussian basis set
expansion approach}
\label{sec_ecg}
To numerically solve the time-independent 
Schr\"odinger 
equation for the Hamiltonian 
given in  
Eq.~(\ref{general Hamiltonian}), 
we employ the 
finite-range
two-body model potential defined 
in 
Eq.~(\ref{gauss potential}). 
We expand the wave function in terms of explicitly
correlated Gaussian basis functions~\cite{suzuki,RMPreview}, which depend
on a set of non-linear variational parameters.
These non-linear parameters are optimized semi-stochastically~\cite{svm}
by minimizing the energy of the state of interest.
Since the
basis functions are not linearly independent,
the eigen energies 
are
obtained by 
solving a generalized eigen value problem 
that involves the Hamiltonian matrix and the overlap 
matrix~\cite{suzuki,RMPreview}. 

In the cold atom context, 
explicitly correlated Gaussian basis sets have been applied
extensively to harmonically trapped 
few-body systems~\cite{sorenson,stec07c,blum07,blume1,blume2,blum11,debraj12}. 
However, this approach has not yet been extended to 
cold atom systems with periodic boundary conditions~\cite{varga}.
To treat periodic systems,  
we imagine that the 
full three-dimensional 
space
is divided into an infinite number of 
cubic
boxes of length $L$. 
We place the
$N$ particles in the 
``center box''. 
The center box defines
our system of interest.
We then 
imagine that the particles in the center box are 
copied to all other boxes, i.e.,
we shift all position vectors $\mathbf{x}_a$
($a=1,\cdots,N$) by $(L b_a^{(1)},L b_a^{(2)},L b_a^{(3)})$,
where the $b_{a}^{(j)}$ 
take the 
values $\cdots,-2,-1,0,1,\cdots$.
Correspondingly, we enforce the periodicity of
the basis functions by explicitly summing over all
possible $b_a^{(j)}$.
The explicit functional form of the basis functions
as well as compact expressions for 
the Hamiltonian matrix 
element
and the overlap matrix 
element 
are given in the Appendix.  

In practice, we can only treat a
finite and not an infinite
number of boxes.
Our calculations reported in Sec.~\ref{sec_results}
employ $9^3$ boxes 
for the $(1,1)$
and
$(2,1)$ systems,  
and 
$7^3$ boxes for 
the $(2,2)$ and $(3,1)$ 
systems. 
We estimate that the error caused by using 
a finite and not an infinite number of boxes is of
the order of 
0.0001\% for the $(1,1)$ and $(2,1)$ systems
and of the order of 0.001\% for the $(2,2)$ and $(3,1)$ systems,
respectively. 
For the $(2,1)$, $(2,2)$
and $(3,1)$
systems, 
this error is significantly smaller than 
the basis set extrapolation error and the error arising from extrapolating
the finite-range energies to the zero-range limit 
(see Sec.~\ref{sec_results} for details).

One of the challenges in constructing numerically
tractable basis sets applicable to cold atom systems is that
the system dynamics depends on the
range $r_0$ of the underlying two-body potential
as well as
the box length $L$,
where
$r_0 \ll L$.
As we will demonstrate in Sec.~\ref{sec_results}, our
basis functions are flexible enough to describe 
short-range
correlations 
that occur at the length scale of $r_0$
and long-range correlations
that occur at the length scale of $L$.
Our scheme to optimize the non-linear parameters 
roughly follows that discussed in Refs.~\cite{suzuki,debraj12}.
In particular, we construct
separate basis sets for each state of interest.
When optimizing highly excited states, we first perform a rough 
minimization of the energy
of all lower-lying states and then use the majority
of the basis functions to minimize the energy of the state of interest.
The calculations 
reported
in  
Sec.~\ref{sec_results}
use of the order of
$N_b=500$, where
$N_b$ is the number of unsymmetrized basis functions.

\section{Results}
\label{sec_results}
This section discusses the energies of the $(1,1)$,
$(2,1)$, $(2,2)$ and $(3,1)$ systems obtained by the
explicitly correlated Gaussian basis set expansion approach.
Throughout, we refer to these energies as ECG energies.

\begin{figure}
\centering
\includegraphics[width=0.45\textwidth]{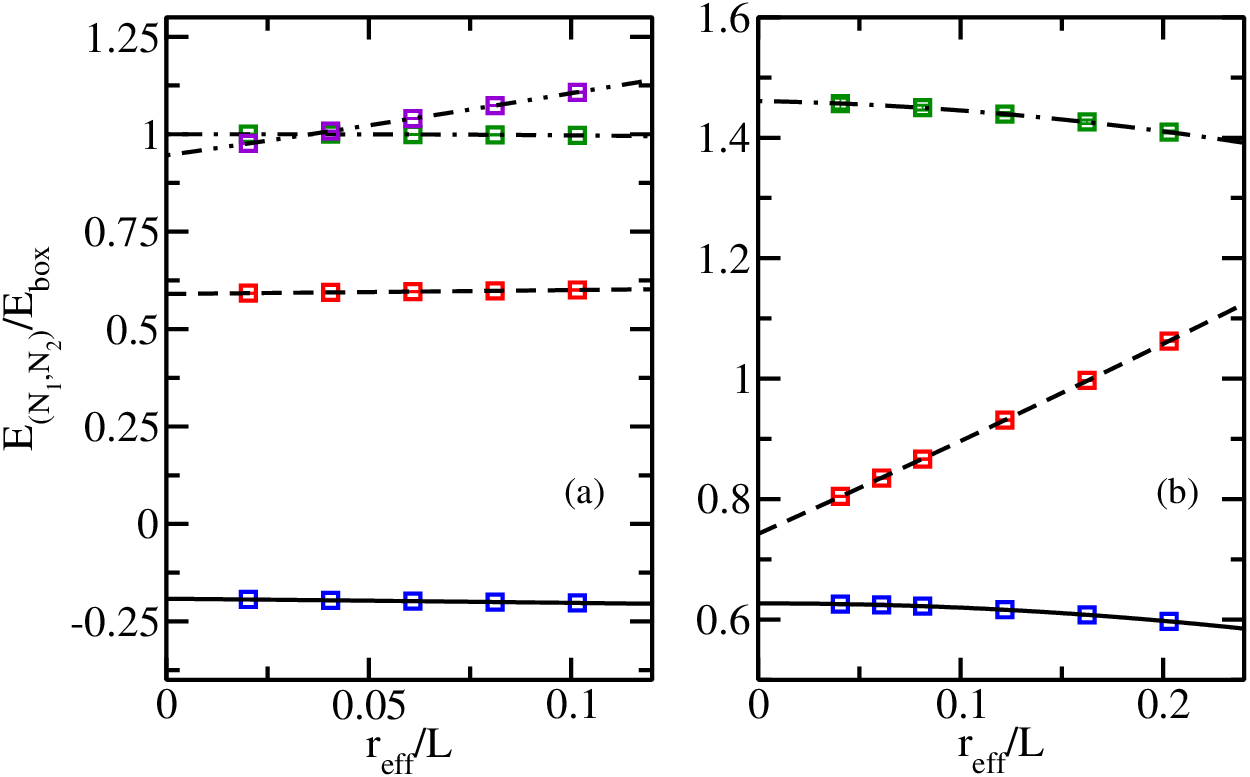}
\caption{(Color online)
(a) The four lowest $(1,1)$ states 
and (b) the three lowest $(2,1)$ states 
at unitarity as a function of 
the effective range
$r_{\text{eff}}$. 
For the Gaussian potential
$V_{\text{g}}$, 
we find
$r_{\text{eff}}
\approx 2.03r_0$. 
Squares with error bars show the  
ECG energies 
extrapolated to the $N_b\rightarrow\infty$ limit
(the error bars are hardly visible on the scale shown). 
The lines show 
fits.}
\label{fig_unitarity11}
\end{figure}

Figure~\ref{fig_unitarity11}(a) 
shows the energies 
of the four lowest  
states
of the $(1,1)$ system 
at unitarity as a function of 
the effective range
$r_{\text{eff}}$.
The effective range is defined through the low-energy expansion
of the two-body free-space $s$-wave 
scattering length~\cite{newtonbook}.
The lowest
state shown in  
Fig.~\ref{fig_unitarity11}(a)
is one-fold degenerate and has vanishing 
momentum 
$\hbar \mathbf{K}$.
The second state has $\hbar |\mathbf{K}|=2 \pi \hbar/L$
and is six-fold degenerate.
The third and fourth states cross at
$r_{\text{eff}}/L \approx 0.04$.
The state that is essentially unaffected
by the interactions 
[dash-dotted line
in Fig.~\ref{fig_unitarity11}(a)] 
is six-fold degenerate; 
in the weakly-attractive
regime,
this state is characterized by $\Delta E_{n,q}=0$.
The state that is more strongly affected by the interactions
[dash-dot-dotted line
in Fig.~\ref{fig_unitarity11}(a)] 
is
one-fold degenerate and has vanishing 
momentum 
$\hbar \mathbf{K}$.

Table~\ref{table_unitarity11} lists the
$(1,1)$
energies for different $r_0$.
\begin{table}
\centering
\caption{Energies of 
the
four lowest states
of the $(1,1)$ 
system at unitarity for different $r_0$.  
The fourth column 
reports
the lowest ECG 
energy for each $r_0$ 
(i.e., the energy for the largest basis set considered). 
The third column  
reports the  
energies extrapolated
to the infinite basis set limit,
i.e., for $N_b\rightarrow\infty$. 
The $r_0=0$ energies are obtained by 
extrapolating the finite-range energies to the zero-range limit.
The error bar for the $r_0=0$ 
energy reported in the third column is obtained by taking the
difference between the $r_0=0$ energies reported in columns 
three and four.
Column five reports the magnitude of the wave vector $|\mathbf{K}|$.}
\label{table_unitarity11}
\centering
\begin{tabular}{cc cc c}
\hline
\hline
state & $r_0/L$ &  $E/E_{\text{box}}$ &  $E/E_{\text{box}}$ & 
$|\mathbf{K}| / \left(\frac{2 \pi}{L} \right)$ \\
 &                   &    $N_b \rightarrow \infty$ & largest $N_b$ & \\
\hline
1   & 0.05   & $-0.20259$   &  $-0.20259$&0\\
    & 0.04   & $-0.20025$  &  $-0.20025$&\\
    & 0.03   & $-0.19801$  &  $-0.19801$&\\
    & 0.02  &  $-0.19586$  &  $-0.19585$&\\
    & 0.01  &  $-0.19379$  &  $-0.19378$&\\                  
\cline{2-4}
    & 0       & $-0.19182(2)$  &  $-0.19180$&\\
\hline                            
2  & 0.05   &  0.60026 &  0.60030&1\\
    & 0.04   &  0.59816  &  0.59820&\\
    & 0.03   &  0.59610  &  0.59623&\\
    & 0.02   &  0.59410  &  0.59437&\\
    & 0.01   &  0.59211  &  0.59245&\\                  
\cline{2-4}
    & 0        &  0.59019(50)    &  0.59069&\\
\hline
3  & 0.05   &   1.10766  &  1.10767 &0\\
    & 0.04   &   1.07358  &  1.07359 &\\
    & 0.03   &   1.04024  &  1.04026 &\\
    & 0.02   &   1.00778  &  1.00784 &\\
    & 0.01   &   0.97623  &  0.97634 &\\
\cline{2-4}
    & 0        &   0.94572(16)  &  0.94588 &\\
\hline                            
4  & 0.05     &  0.99689  &  0.99692&1\\
     & 0.04    &   0.99840  &  0.99842&\\
    & 0.03     &  0.99931  &  0.99935&\\
    & 0.02     &  0.99979  &  0.99982&\\
    & 0.01     &  0.99997  &  1.00000&\\                  
\cline{2-4}
    & 0        &  1.00000(2)    &  1.00002&\\    
\hline
\hline
\end{tabular}
\end{table} 
The fourth column reports the energies for the largest basis set considered;
according to the
variational principle~\cite{suzuki,RMPreview}, these ECG energies
provide upper bounds 
to the exact eigen energies.
The third column reports the 
energies obtained by extrapolating 
the ECG
energies to the infinite basis set limit.
To extrapolate the 
finite-range
energies to the 
zero-range
limit,
we perform separate three or four parameter fits to the energies
listed in the third and fourth columns of Table~\ref{table_unitarity11}. 
The $N_b \rightarrow \infty$ energies, extrapolated to the 
zero-range limit, are our best estimates for the 
zero-range energies. The associated error bars 
(see Table~\ref{table_unitarity11}) are obtained by taking
the difference between the 
extrapolated zero-range energies reported in columns
three and four.
Imposing the Bethe-Peierls boundary condition
for zero-range interactions 
on the  two-body wave function, the zero-range energies
for states
with $\mathbf{K}=0$ can 
be found with very high accuracy~\cite{luscher1,luscher2}
(in nuclear physics, the resulting implicit eigen
equation is known as ``L\"uscher formula'').
For the two lowest $\mathbf{K}=0$ levels one finds
$E=-0.19180 E_{\text{box}}$ and
$E=0.94579E_{\text{box}}$, respectively.
Our zero-range energies
[$E=-0.19182(2)E_{\text{box}}$
and 
$E=0.94572(16)E_{\text{box}}$,
see Table~\ref{table_unitarity11}]
agree with the exact energies
within error bars.
Our energy of $E=0.59019(50)E_{\text{box}}$
for the lowest state with $\hbar |\mathbf{K}|=2  \pi \hbar  / L$
agrees with the value of
$E=0.5902 E_{\text{box}}$ obtained by Werner and Castin~\cite{castin12}.

Figure~\ref{fig_unitarity11}(b) and 
Table~\ref{table_unitarity21}
summarize our results for the $(2,1)$ system at unitarity.
\begin{figure}
\centering
\includegraphics[width=0.3\textwidth]{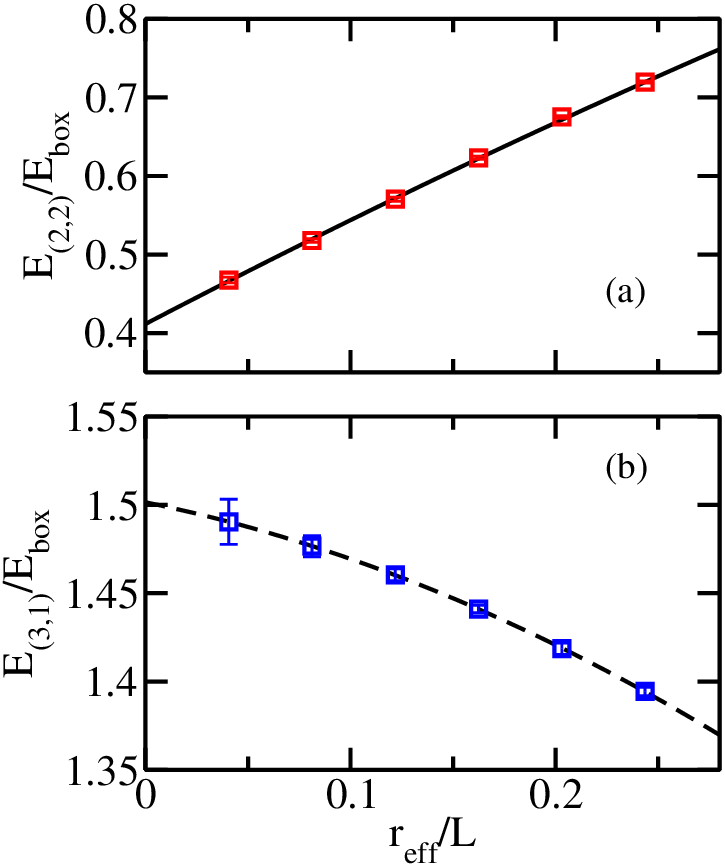}
\caption{(Color online)
Energy of the lowest state of (a) the $(2,2)$ system 
and (b) the $(3,1)$ system 
at unitarity as a function of $r_{\text{eff}}$.
See caption of Fig.~\ref{fig_unitarity11} for details. 
}
\label{fig_unitarity22and31}
\end{figure}
The lowest state of the $(2,1)$ system at unitarity
is six-fold degenerate and
has $\hbar | \mathbf{K}| = 2 \pi \hbar / L$,
while the second and third states
are one-fold and three-fold degenerate,
respectively, and have
$\hbar | \mathbf{K}| = 0$.
\begin{table}
\centering
\caption{Energies of 
the three lowest levels of the $(2,1)$ 
system at unitarity for different $r_0$.  See 
caption of Table~\ref{table_unitarity11}
for details.
For the first excited state, the difference between
the zero-range energies $E/E_{\text{box}}$ calculated for infinite
and finite $N_b$ is very small; we estimate that this difference
underestimates the ``true'' error bar.
}
\label{table_unitarity21}
\centering
\begin{tabular}{ccccc}
\hline
\hline
state & $r_0$/$L$ &  $E/E_{\text{box}}$ & $E/E_{\text{box}}$&
$|\mathbf{K}|/\left( \frac{2 \pi}{L} \right)$\\
&                   &    $N_b \rightarrow \infty$ & largest $N_b$&  \\
\hline
1  & 0.1    &  0.5971  &  0.5973 & 1\\
    & 0.08  &  0.6080  &  0.6085 & \\
    & 0.06  &  0.6164  &  0.6168 & \\
    & 0.04  &  0.6223  &  0.6227 & \\
    & 0.03  &  0.6244  &  0.6256 & \\
    & 0.02  &  0.6259  &  0.6270 & \\                  
    & 0.01  &  0.6274  &  0.6295 & \\
\cline{2-4}
    & 0       & 0.6282(30)  &  0.6312 & \\
\hline                            
2  & 0.1 &  1.0625  &  1.0626&0\\
    & 0.08   &  0.9971  &  0.9972&\\
    & 0.06   &  0.9312  &  0.9314&\\
    & 0.04   &  0.8665  &  0.8668&\\
    & 0.03   &  0.8352  &  0.8356&\\
    & 0.02   &  0.8047  &  0.8051&\\                  
\cline{2-4}
    & 0        &  0.7424(6)    &  0.7430&\\
\hline                            
3  & 0.1     &  1.4095  &  1.4112&0\\
    & 0.08   &  1.4256  &  1.4287&\\
    & 0.06   &  1.4392  &  1.4421&\\
    & 0.04   &  1.4500  &  1.4540&\\
    & 0.02   &  1.4567  &  1.4657&\\                  
\cline{2-4}
    & 0        &  1.4609(132)    &  1.4741&\\    
\hline
\hline
\end{tabular}
\end{table} 
The $(2,1)$ energies at unitarity have previously been 
determined by a variety of methods,
including a continuum
Green's function approach~\cite{castin}
and lattice Monte Carlo techniques~\cite{kaplan,deanlee}.
While states with vanishing momentum have been
considered frequently,
we
are aware of only one study that considered 
states
with finite total momentum~\cite{olivier}.
Our energies for the second
and third states agree, within error bars, with the
literature values~\cite{castin}. 
Our estimate for the $(2,1)$ zero-range ground state 
energy at unitarity is $E=0.6282(30) E_{\text{box}}$.
The fact that the ground state has finite total momentum is a direct
consequence of the anti-symmetry requirement of the wave function
under the interchange of the two identical fermions.
This is analogous to the harmonically trapped $(2,1)$ system
at unitarity with
zero-range interactions, which is characterized by a total orbital angular
momentum of $L=1$~\cite{blume1,kest07,stet07}.

Figure~\ref{fig_unitarity22and31}  and 
Table~\ref{table_unitarity22and31} 
summarize our results for the lowest state of 
the $(2,2)$ and $(3,1)$ systems at unitarity. 
These states
are one-fold and three-fold degenerate, respectively,
and have  $\hbar | \mathbf{K}| = 0$. 
\begin{table}
\centering
\caption{Energies of 
the lowest state of the $(2,2)$ and
$(3,1)$ 
systems at unitarity for different $r_0$.  See 
caption of Table~\ref{table_unitarity11}
for details.
}
\label{table_unitarity22and31}
\centering
\begin{tabular}{ccccc}
\hline
\hline
$(N_1, N_2)$ & $r_0$/$L$ &  $E/E_{\text{box}}$ & $E/E_{\text{box}}$&
$|\mathbf{K}| / \left( \frac{2 \pi}{L} \right)$\\
&                   &    $N_b \rightarrow \infty$ & largest $N_b$  &\\
\hline
(2,2)  & 0.12  &  0.7182  &  0.7196&0\\
         & 0.1    &  0.6735  &  0.6751&\\
         & 0.08  &  0.6232  &  0.6252&\\
         & 0.06  &  0.5704  &  0.5730&\\
         & 0.04  &  0.5178  &  0.5200&\\
         & 0.02  &  0.4675  &  0.4712&\\                  
\cline{2-4}
         & 0       & 0.4116(42)  &  0.4158&\\
\hline                            
(3,1)  & 0.12 &  1.3944  &  1.3968&0\\
         & 0.1   &  1.4186  &  1.4218&\\
         & 0.08 &  1.4410  &  1.4431&\\
         & 0.06 &  1.4604  &  1.4638&\\
         & 0.04 &  1.4766  &  1.4826&\\
         & 0.02 &  1.4904  &  1.5032& \\                  
\cline{2-4}
         & 0      & 1.5014(187) &  1.5201&\\
\hline
\hline
\end{tabular}
\end{table} 
The ground state energy of 
the $(2,2)$ system has been benchmarked previously~\cite{deanlee}.
Reference~\cite{deanlee}
finds
$E=0.422(4) E_{\text{box}}$ and $0.420(4)E_{\text{box}}$ using two different 
lattice representations of the Hamiltonian,
$E= 0.412(18) E_{\text{box}}$ using a Euclidean lattice approach,
and an upper bound of $E=0.424(4) E_{\text{box}}$ 
using the fixed-node diffusion Monte Carlo
approach.
Our extrapolated zero-range energy of
$E= 0.4116(42) E_{\text{box}}$ agrees with these results
within error bars. Note that our error bar is comparable to those
of Ref.~\cite{deanlee}.
For the $(3,1)$ system, we are not aware of any literature results.

\begin{figure}
\centering
\includegraphics[width=0.3\textwidth]{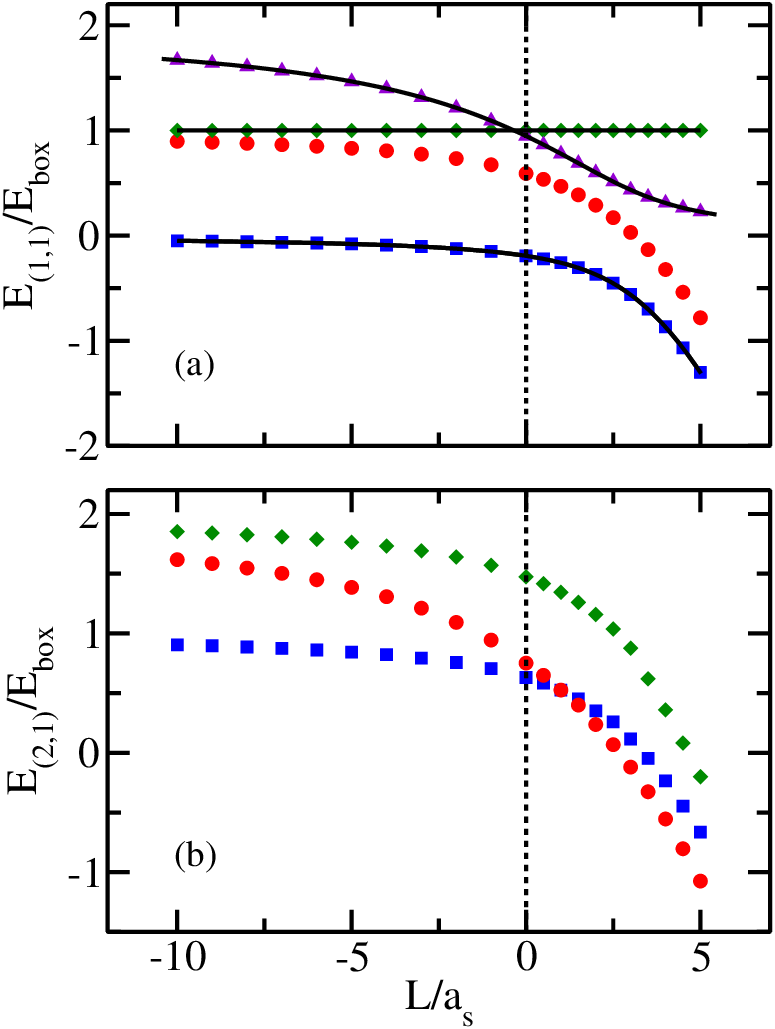}
\caption{(Color online) Zero-range energies of 
(a) the four lowest states 
of the $(1,1)$ system 
and (b) the 
three lowest states of the 
$(2,1)$ system as a function of $L/a_s$. 
The symbols show the 
lowest ECG energies extrapolated to 
the $r_0\rightarrow0$ limit. 
The solid lines in panel (a) for the $\mathbf{K}=0$
states are obtained from L\"uscher's
formula~\protect\cite{luscher1,luscher2}.}
\label{fig_cross11}
\end{figure}

Symbols in Figs.~\ref{fig_cross11}(a) and \ref{fig_cross11}(b)
show the lowest few levels 
of the
$(1,1)$ and $(2,1)$ systems with 
zero-range interactions as a function of $L/a_s$, i.e.,
throughout the BCS to BEC crossover.
The energies are obtained by extrapolating our finite-range
ECG energies to the zero-range limit for each
$a_s/L$.
In the weakly-attractive BCS regime ($a_s <0$ and
$|a_s|/L \ll 1$), our 
extrapolated zero-range energies agree with the perturbative energies
discussed  in
Sec.~\ref{sec_pt}. 
Our $(1,1)$ energies for states with vanishing momentum 
[squares and triangles in Fig.~\ref{fig_cross11}(a)]
are in excellent agreement with the energies obtained 
from L\"uscher's formula [see
solid lines in Fig.~\ref{fig_cross11}(a)]~\cite{luscher1,luscher2}.
In the BEC regime ($a_s>0$ and $a_s/L \ll 1$), 
the energy spectrum contains two types of energy levels,
those where the corresponding states ``contain'' dimers
[e.g., the lowest level in Figs.~\ref{fig_cross11}(a) and 
\ref{fig_cross11}(b)]
and those where the corresponding states are best thought of as
describing an atomic gas [see triangles and diamonds 
in Fig.~\ref{fig_cross11}(a)].
The dimers consist of
fermions that have opposite spin projections.
In the $(1,1)$ system,
states with vanishing and finite momentum can form 
$s$-wave dominated dimers. 
This can be readily understood by realizing that 
the total momentum and the orbital angular momentum are
distinctly different quantities and that states with
finite total momentum contain $s$-wave contributions~\cite{luscher1,luscher2}.
The $(2,1)$ energy spectrum shows a crossing of the two lowest states
at 
$L/a_s \approx 1$. 
The state with
$\hbar | \mathbf{K}| = 2 \pi \hbar /L$ has a lower
energy in the BCS regime
while the state with
$\hbar | \mathbf{K}| = 0$ has a lower
energy in the BEC regime.
This crossing is somewhat similar to the crossing between states with 
finite and vanishing orbital
angular momentum in the harmonically trapped $(2,1)$ 
system~\cite{blume1,kest07,stet07}.

\begin{figure}
\vspace*{1in}
\centering
\includegraphics[width=0.3\textwidth]{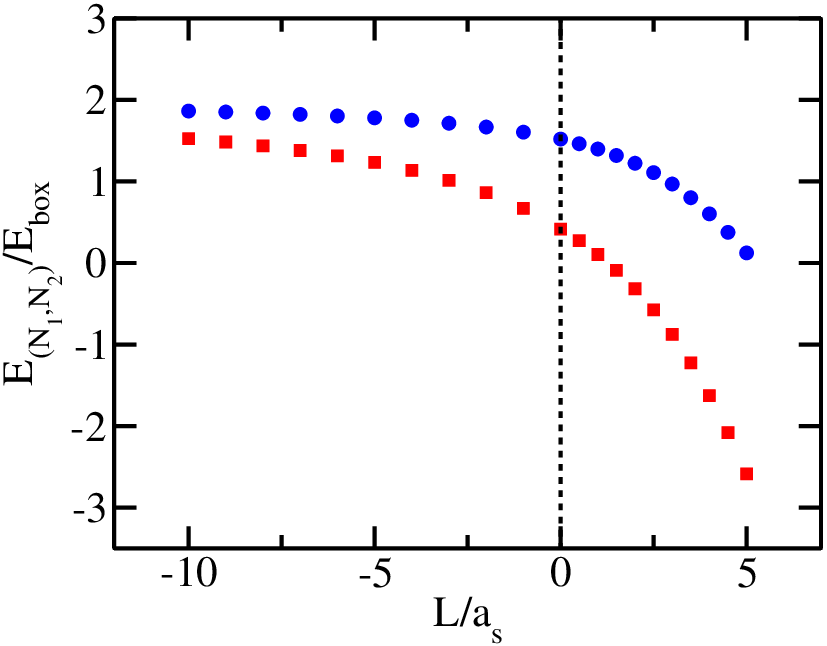}
\caption{(Color online) Zero-range energies of the lowest state
of the $(2,2)$ and $(3,1)$ systems as a function of $L/a_s$. 
Squares and circles show the lowest ECG energy
extrapolated to the $r_0\rightarrow0$ limit
for the $(2,2)$ and $(3,1)$ systems, respectively.}
\label{fig_cross22and31}
\end{figure}

Squares and circles in Fig.~\ref{fig_cross22and31} show the extrapolated 
zero-range energies
of the ground state of the $(2,2)$ system 
and the $(3,1)$ system, respectively, as a 
function of $L/a_s$. 
In the $a_s \rightarrow 0^-$ limit, 
the energies
of the $(2,2)$ and
$(3,1)$ systems agree.
In the $a_s \rightarrow 0^+$ limit, 
in contrast, the energy of the $(2,2)$ 
system is significantly lower
than that of the $(3,1)$ system, 
reflecting the fact that the $(2,2)$ and $(3,1)$
systems
form two dimers and one dimer, respectively.

\section{Conclusion}
\label{sec_sumary}
This paper considered the energetics of small two-component Fermi gases
in a cubic box with periodic boundary conditions.
We treated systems with up to $N=4$ atoms using
first-order perturbation theory and the explicitly correlated
Gaussian basis set expansion approach.
We determined the low-lying states throughout the BCS-BEC 
crossover and carefully analyzed the dependence of the energies on the range
of the underlying two-body potential at unitarity.
Our calculations agree with results reported in the literature
and are expected to serve as benchmarks in the cases where no
other literature values exist.
The method introduced in this paper extends the
application of explicitly correlated Gaussian basis sets,
optimized using the
stochastic variational method, to quantum few-body
problems with periodic boundary conditions. 

\section{Acknowledgement}
Discussions with D. Kaplan, D. Lee and K. Varga, 
which motivated this work, correspondence with
O. Juillet as well as
support by the National
Science Foundation (NSF) through Grant No.
PHY-1205443
are greatfully acknowledged.
This work used the Extreme Science and Engineering
Discovery Environment (XSEDE), which is supported by
NSF grant number OCI-1053575, and the
WSU HPC.
This work was additionally supported by the 
NSF through a grant for the Institute 
for Theoretical Atomic, Molecular and Optical Physics 
at Harvard University and
Smithsonian Astrophysical Observatory.

\onecolumngrid
\appendix
\section{Explicitly correlated Gaussian basis functions
for systems with periodic boundary conditions}

This appendix introduces 
explicitly correlated Gaussian basis functions 
that obey periodic boundary conditions
and derives analytic expressions
for the overlap, kinetic energy 
and interaction matrix elements.
Throughout this appendix, we do not impose symmetry constraints.
The proper symmetry
of the basis functions can be enforced
following Sec.~2.3 of Ref.~\cite{suzuki}.

The system Hamiltonian $H$ is the
sum of the kinetic energy $H_0$, Eq.~(\ref{NI Hamiltonian}), 
and the particle-particle
interactions $V_{\text{int}}$, Eq.~(\ref{interaction}).
As discussed in Sec.~\ref{sec_ecg},
we imagine that the full three-dimensional space
is divided into an infinite number of cubic boxes of length $L$. 
The lower left corner of the center box, labeled by $\mathbf{0}$,
is located at the origin.
Unlike particles are interacting through the finite-range two-body 
Gaussian potential $V_{\text{g}}$, Eq.~(\ref{gauss potential}).
To account for the periodicity of the system, we write
\begin{eqnarray} \label{3D interaction} 
V_{\text{int}}^{\text{pbc}}&=&
\sum_{a=1}^{N_1} \sum_{b=N_1+1}^{N} \sum_{\mathbf{q}} 
V_{\text{g}}(\mathbf{x}_{ab}-L\mathbf{q}),
\end{eqnarray}
where
$\mathbf{q}^{T}=(q^{(1)}, q^{(2)},q^{(3)})$
denotes a three-component vector with
$q^{(i)}= \cdots, -2, -1, 0, 1, 2, \cdots$.
The sum over $\mathbf{q}$ in Eq.~(\ref{3D interaction}) ensures that
we are not only considering interactions between pairs 
of particles of opposite spin in box $\mathbf{0}$
but also interactions of particles in box $\mathbf{0}$ with particles
of opposite 
spin located in other boxes.
A key advantage of the Gaussian potential $V_{\text{g}}$ is that it 
factorizes,
\begin{eqnarray}
V_{\text{int}}^{\text{pbc}}=
\sum_{a=1}^{N_1} \sum_{b=N_1+1}^{N} 
U_0 \prod_{i=1}^3
V_{\text{g}}^{{\text{pbc}},(i)}({x}_{ab}^{(i)}-L{q}^{(i)}),
\end{eqnarray}
where
$V_{\text{g}}^{\text{pbc},(i)}(x_{ab}^{(i)})=
\sum_{\mathbf{q}} \exp[-(x_{ab}^{(i)})^2/(2 r_0^2)]$.

\subsection{Basis functions}
\label{app_basis}
We first focus on the $i$th spatial dimension.
To construct basis functions $\Psi^{(i)}$ for the $i$th
spatial dimension, we introduce a 
``single particle'' $N\times N$ matrix $B^{(i)}$,
a ``two-body'' $N\times N$ matrix $A^{(i)}$ 
and a displacement vector $\mathbf{s}^{(i)}$,
$\mathbf{s}^{(i)}=(s_1^{(i)}, \cdots, s_N^{(i)})$,
and consider
the unsymmetrized and non-periodic function 
$\Psi_{\text{np}}^{(i)}$,
\begin{equation} \label{1D basis}
\Psi_{\text{np}}^{(i)} (A^{(i)}, B^{(i)}, \mathbf{s}^{(i)}, \mathbf{x}^{(i)})
=\exp \left[ -\frac{1}{2} ( \mathbf{x}^{(i)} )^{T} 
A^{(i)} \mathbf{x}^{(i)}
-\frac{1}{2}(\mathbf{x}^{(i)}-\mathbf{s}^{(i)})^{T} B^{(i)} (\mathbf{x}^{(i)}-\mathbf{s}^{(i)})\right] ,
\end{equation}
where
$( \mathbf{x}^{(i)} )^T=(x_1^{(i)}, \cdots, x_N^{(i)})$.
The function $\Psi_{\text{np}}^{(i)}$ can alternatively be written in terms of 
``single particle'' Gaussian widths $d_a^{(i)}$ and
``two-body'' Gaussian widths $d_{ab}^{(i)}$,
\begin{equation}
\Psi_{\text{np}}^{(i)}(d_{12}, \cdots, d_{N-1, N}, d_1, \cdots, d_N, 
\mathbf{s}^{(i)}, \mathbf{x}^{(i)})=
\exp\left[ -\sum_{a=1}^{N} \sum_{b=a+1}^{N} 
\frac{(x_a^{(i)}-x_b^{(i)})^2}{2(d_{ab}^{(i)})^2} \right]  
\exp\left[-\sum_{a=1}^{N} \frac{(x_a^{(i)}-s_a^{(i)})^2}{2(d_a^{(i)})^2}\right].
\end{equation} 
The diagonal elements of the matrix $B^{(i)}$ are related to 
the $d_a^{(i)}$ by
$B_{aa}^{(i)}=1/(d_a^{(i)})^2$.
The off-diagonal elements of $B^{(i)}$ are zero.
$A^{(i)}$ is a symmetric matrix constructed from the
$N(N-1)/2$ independent
Gaussian widths $d_{ab}^{(i)}$.
Transforming from relative distance coordinates to single-particle
coordinates, we have
$A_{ab}^{(i)}=-(d_{ab}^{(i)})^{-2}$
for $a\neq b$
and
$A_{aa}^{(i)}=\sum_{b=1, b \neq a}^{N} (d_{ab}^{(i)})^{-2}$.

The function $\Psi_{\text{np}}^{(i)}$ introduced in Eq.~(\ref{1D basis})
does not obey periodic boundary conditions. To enforce
periodic boundary conditions, we introduce a sum over the vector
$\mathbf{b}^{(i)}$,
\begin{eqnarray}
\Psi^{(i)}(A^{(i)}, B^{(i)}, \mathbf{s}^{(i)}, \mathbf{x}^{(i)}) 
=
\sum_{\mathbf{b}^{(i)}}
\Psi_{\text{np}}^{(i)} (A^{(i)}, B^{(i)}, \mathbf{s}^{(i)}, 
\mathbf{x}^{(i)}-L\mathbf{b}^{(i)}) ,
\end{eqnarray}
where
$( \mathbf{b}^{(i)} )^{T}=(b_1, b_2, \cdots, b_N)$
with
$b_j^{(i)}= \cdots , -2, -1, 0 ,1, 2, \cdots$.
It can be readily checked that
$\Psi^{(i)} (A^{(i)}, B^{(i)}, \mathbf{s}^{(i)}, \mathbf{x}^{(i)}-\mathbf{t})$,
where $\mathbf{t}$ is a $N$-component vector with a single non-zero
element, $\mathbf{t}^T=(0, \cdots, 0, L, 0, \cdots, 0)$,
equals $\Psi^{(i)} (A^{(i)}, B^{(i)}, \mathbf{s}^{(i)}, \mathbf{x}^{(i)})$,
that is, $\Psi^{(i)}$ obeys periodic boundary conditions.
The three-dimensional unsymmetrized basis function $\Psi_{\text{3D}}$ is 
simply the product of the basis functions in the $x$-, $y$- and $z$-directions,
i.e., $\Psi_{\text{3D}}= \prod_{i=1}^3 \Psi^{(i)}$.

\subsection{Overlap matrix element}
\label{app_overlap}
The overlap between the basis functions $\Psi_{\text{3D}}$ and
$\Psi_{\text{3D}}'$
is
\begin{eqnarray}
\label{eq_overlap1}
\langle \Psi_{\text{3D}} | \Psi_{\text{3D}}' \rangle
=
\prod_{i=1}^{3} 
\langle \Psi^{(i)} | \Psi'^{(i)} \rangle =
\prod_{i=1}^{3} 
\left[
\int_0^L \cdots \int_0^L
 \Psi^{(i)} (A^{(i)}, B^{(i)}, \mathbf{s}^{(i)}, \mathbf{x}^{(i)}) 
\Psi^{(i)} (A'^{(i)}, B'^{(i)}, \mathbf{s}'^{(i)}, \mathbf{x}^{(i)})
 d {\mathbf{x}}^{(i)} 
\right]
.
\end{eqnarray}
In the following, we focus on the overlap matrix element
for the $i$th dimension.
To perform the integration analytically, we shall
change 
the integration limits from $[0, L]$ to $[-\infty, \infty]$. As a first step,
we shift the spatial coordinates by 
defining 
$\mathbf{x}_{\text{new}}^{(i)}=\mathbf{x}^{(i)}-L\mathbf{b}'^{(i)}$
and then renaming $\mathbf{x}_{\text{new}}^{(i)}$ as $\mathbf{x}^{(i)}$
for convenience,
\begin{eqnarray}
\langle \Psi^{(i)} | \Psi'^{(i)} \rangle &=&
\sum_{\mathbf{b}^{(i)}} \sum_{\mathbf{b}'^{(i)}}
\int_{-Lb_1'^{(i)}}^{L-Lb_1'^{(i)}} \cdots \int_{-Lb_N'^{(i)}}^{L-Lb_N'^{(i)}}
\exp \bigg[ -\frac{1}{2} (\mathbf{x}^{(i)}+L\mathbf{b}'^{(i)}-L\mathbf{b}^{(i)})^T
A^{(i)} (\mathbf{x}^{(i)}+L\mathbf{b}'^{(i)}-L\mathbf{b}^{(i)}) 
 \nonumber \\
&& -\frac{1}{2}(\mathbf{x}^{(i)}-\mathbf{s}^{(i)}
+L\mathbf{b}'^{(i)}-L\mathbf{b}^{(i)})^{T} 
B^{(i)} (\mathbf{x}^{(i)}-\mathbf{s}^{(i)}
+L\mathbf{b}'^{(i)}-L\mathbf{b}^{(i)})\biggr] \times
\nonumber \\
&&\exp \left[ -\frac{1}{2} (\mathbf{x}^{(i)})^T 
A'^{(i)} (\mathbf{x}^{(i)})
-\frac{1}{2}(\mathbf{x}^{(i)}-\mathbf{s}'^{(i)})^{T} 
B'^{(i)} (\mathbf{x}^{(i)}-\mathbf{s}'^{(i)})\right]
d \mathbf{x}^{(i)}.
\end{eqnarray}
Next, we 
replace $\mathbf{b}^{(i)}-\mathbf{b}'^{(i)}$ by
$\Delta \mathbf{b}^{(i)}$
and replace the sum over  $\mathbf{b}^{(i)}$ by a sum over 
$\Delta \mathbf{b}^{(i)}$,
\begin{eqnarray}
\langle \Psi^{(i)} | \Psi'^{(i)} \rangle &=&
\sum_{ \Delta \mathbf{b}^{(i)}} \sum_{\mathbf{b}'^{(i)}}
\int_{-Lb_1'^{(i)}}^{L-Lb_1'^{(i)}} \cdots \int_{-Lb_N'^{(i)}}^{L-Lb_N'^{(i)}}
\exp \biggl[ -\frac{1}{2} (\mathbf{x}^{(i)}-L\Delta \mathbf{b}^{(i)})^T
A^{(i)} (\mathbf{x}^{(i)}-L\Delta \mathbf{b}^{(i)}) \nonumber \\
&&-\frac{1}{2}(\mathbf{x}^{(i)}-\mathbf{s}^{(i)}-L\Delta \mathbf{b}^{(i)})^{T} 
B^{(i)} (\mathbf{x}^{(i)}-\mathbf{s}^{(i)}-L\Delta \mathbf{b}^{(i)}) \biggr]
\times \nonumber \\
&&     
\exp \left[ -\frac{1}{2} (\mathbf{x}^{(i)})^T 
A'^{(i)} (\mathbf{x}^{(i)})
-\frac{1}{2}(\mathbf{x}^{(i)}-\mathbf{s}'^{(i)})^{T} 
B'^{(i)} (\mathbf{x}^{(i)}-\mathbf{s}'^{(i)})\right]
d \mathbf{x}^{(i)} .
\end{eqnarray}
Since the integrand is independent of $\mathbf{b}'^{(i)}$, the sum over
$\mathbf{b}'^{(i)}$ 
changes the integration limits of the $N$ integrals
to $[-\infty, \infty]$.
Renaming $\Delta \mathbf{b}^{(i)}$ as $\mathbf{b}^{(i)}$ for 
convenience, we find
\begin{eqnarray} \label{1D overlap middle}
\langle \Psi^{(i)} | \Psi'^{(i)} \rangle &=&
\sum_{\mathbf{b}^{(i)}}
\int_{-\infty}^{\infty} \cdots \int_{-\infty}^{\infty}
\exp \biggl[ -\frac{1}{2}(\mathbf{x}^{(i)}-L\mathbf{b}^{(i)})^T 
A^{(i)} (\mathbf{x}^{(i)}-L\mathbf{b}^{(i)}) 
\nonumber \\
&&
-\frac{1}{2}(\mathbf{x}^{(i)}-\mathbf{s}^{(i)}-L\mathbf{b}^{(i)})^{T} 
B^{(i)} (\mathbf{x}^{(i)}-\mathbf{s}^{(i)}- L\mathbf{b}^{(i)})\biggr]
\times
\nonumber \\
&&   
\exp \left[ -\frac{1}{2} (\mathbf{x}^{(i)})^T  
A'^{(i)} (\mathbf{x}^{(i)})
-\frac{1}{2}(\mathbf{x}^{(i)}-\mathbf{s}'^{(i)})^{T} 
B'^{(i)} (\mathbf{x}^{(i)}-\mathbf{s}'^{(i)})\right]
d \mathbf{x}^{(i)} .
\end{eqnarray}
In going from Eq.~(\ref{eq_overlap1}) to Eq.~(\ref{1D overlap middle}), 
we have transformed the integrals over box $\mathbf{0}$ to integrals over
all space.

Pulling $\mathbf{x}^{(i)}$-independent terms out of the integrals,
Eq.~(\ref{1D overlap middle}) becomes
\begin{eqnarray} \label{1D overlap middle 2}
\langle \Psi^{(i)} | \Psi'^{(i)} \rangle =
\sum_{\mathbf{b}^{(i)}}
{\cal{C}}^{(i)}(A^{(i)},B^{(i)},\mathbf{s}^{(i)},\mathbf{b}^{(i)})
\int_{-\infty}^{\infty} ... \int_{-\infty}^{\infty}
&&
g \left( A^{(i)} (L\mathbf{b}^{(i)})+B^{(i)} (L\mathbf{b}^{(i)}+\mathbf{s}^{(i)});
A^{(i)}+B^{(i)}, \mathbf{x}^{(i)} \right) \times
\nonumber \\
&&g \left( B'^{(i)} \mathbf{s}'^{(i)}; A'^{(i)}+B'^{(i)}, 
\mathbf{x}^{(i)} \right)d \mathbf{x}^{(i)},
\end{eqnarray}
where
\begin{eqnarray}
{\cal{C}}^{(i)}(A^{(i)},B^{(i)},\mathbf{s}^{(i)},\mathbf{b}^{(i)})
=\\ \nonumber
\exp \biggl[
-\frac{1}{2} (L\mathbf{b}^{(i)})^T A^{(i)} (L\mathbf{b}^{(i)})
-\frac{1}{2} (L\mathbf{b}^{(i)}+\mathbf{s}^{(i)})^T B^{(i)}
 (L\mathbf{b}^{(i)}+\mathbf{s}^{(i)})
-\frac{1}{2} (\mathbf{s}'^{(i)})^T B'^{(i)} \mathbf{s}'^{(i)}
\biggr]
\end{eqnarray}
and
\begin{eqnarray}
g(\mathbf{h}; D, \mathbf{x})=
\exp \left( -\frac{1}{2} \mathbf{x}^{T} D \mathbf{x}
+\mathbf{h}^{T} \mathbf{x} \right)
\end{eqnarray}
is the generating function defined in Eq.~(6.19) of Ref.~\cite{suzuki}.
Using Eqs.~(7.22) and (7.23) of Ref.~\cite{suzuki}, we find
\begin{eqnarray}
\int_{-\infty}^{\infty} ... \int_{-\infty}^{\infty}
g(\mathbf{h}; D, \mathbf{x}) g(\mathbf{h}'; D', \mathbf{x})
d \mathbf{x} 
=\left( \frac{(2 \pi)^N}{\det C} \right)^{1/2}
\exp \left(\frac{1}{2} \mathbf{v}^{T} C^{-1} \mathbf{v} \right)
\end{eqnarray}
with
$C=D+D'$
and
$\mathbf{v}=\mathbf{h}+\mathbf{h}'$.
Substituting 
$D=A^{(i)}+B^{(i)}$, $D'=A'^{(i)}+B'^{(i)}$,
$\mathbf{h}=A^{(i)} (L\mathbf{b}^{(i)})
+B^{(i)} (L\mathbf{b}^{(i)}+\mathbf{s}^{(i)})$
and
$\mathbf{h}'=B'^{(i)} \mathbf{s}'^{(i)}$,
we find
\begin{eqnarray} \label{1D overlap result}
\langle \Psi^{(i)} | \Psi'^{(i)} \rangle =
\sum_{\mathbf{b}^{(i)}}
{\cal{C}}^{(i)}(A^{(i)},B^{(i)},\mathbf{s}^{(i)},\mathbf{b}^{(i)})
\left( \frac{(2 \pi)^N}{\det \left(C^{(i)}\right)} \right)^{1/2}
\exp \biggl[
\frac{1}{2} (\mathbf{v}^{(i)})^{T} (C^{(i)})^{-1} \mathbf{v}^{(i)}
\biggr],
\end{eqnarray}
where
\begin{eqnarray} \label{C definition}
C^{(i)}=A^{(i)}+B^{(i)}+A'^{(i)}+B'^{(i)}
\end{eqnarray}
and
\begin{eqnarray} \label{1D overlap end}
\mathbf{v}^{(i)}=A^{(i)} (L\mathbf{b}^{(i)}) + 
B^{(i)} (L\mathbf{b}^{(i)}+\mathbf{s}^{(i)})
+B'^{(i)} \mathbf{s}'^{(i)} .
\end{eqnarray}

\subsection{Kinetic energy matrix element}
\label{app_kinetic}
The kinetic energy matrix element is given by
\begin{eqnarray}
\langle \Psi_{\text{3D}} |H_0| \Psi_{\text{3D}}' \rangle
=
\sum_{i=1}^3 \left[ \langle \Psi^{(i)} |H_0^{(i)}| \Psi'^{(i)} \rangle
\left( \prod_{j=1,j \ne i}^3  \langle \Psi^{(j)} | \Psi'^{(j)} \rangle \right)
\right]
,
\end{eqnarray}
where $\langle \Psi^{(i)} | \Psi'^{(i)} \rangle$ is given in 
Eq.~(\ref{1D overlap result})
and where 
$H_0^{(i)} = \sum_{a=1}^N \frac{-\hbar^2}{2m} \frac{\partial^2}{\partial (x_a^{(i)})^2}$.
Thus we only need to evaluate the kinetic energy matrix element for the
$i$th dimension.
Following steps similar to those detailed in Sec.~\ref{app_overlap}, we find
\begin{eqnarray}
\langle \Psi ^{(i)}
 | H_0^{(i)}| \Psi'^{(i)} \rangle
=
\sum_{\mathbf{b}^{(i)}}
\frac{\hbar^2}{2} 
\left[  \Tr  \left( (A^{(i)}+B^{(i)})(C^{(i)})^{-1}(A'^{(i)}+B'^{(i)})
\Lambda \right)
-(\mathbf{y}^{(i)})^{T} \Lambda \mathbf{y}^{(i)} \right]
\langle \Psi ^{(i)}
 |\Psi'^{(i)} \rangle,
 \end{eqnarray}
where
\begin{equation}
\mathbf{y}^{(i)}=(A'^{(i)}+B'^{(i)})(C^{(i)})^{-1} 
\left[ B^{(i)}(L\mathbf{b}^{(i)}+\mathbf{s}^{(i)})
+A^{(i)}(L\mathbf{b}^{(i)}) \right]
-(A^{(i)}+B^{(i)}) (C^{(i)})^{-1} ( B'^{(i)} \mathbf{s}'^{(i)} )
\end{equation}
and
$\Lambda$ is a $N \times N$ diagonal matrix with diagonal elements
$\Lambda_{jj}=1/m_j$;
here, $m_j$ is the mass of the $j$th atom
 (in our case, $m_j=m$).

\subsection{Interaction matrix element}
\label{app_interaction}
The result for the interaction matrix element
$\langle \Psi_{\text{3D}} |V_{\text{int}}^{\text{pbc}}| \Psi_{\text{3D}}' \rangle$
can be readily constructed from
the matrix elements
$\langle \Psi^{(i)} |V_{\text{g}}^{\text{pbc},(i)}| \Psi'^{(i)} \rangle$.
To evaluate $\langle \Psi^{(i)} |V_{\text{g}}^{\text{pbc},(i)}| \Psi'^{(i)} \rangle$,
we define 
$\mathbf{x}_{\text{new}}^{(i)}=\mathbf{x}^{(i)}-L\mathbf{b}'^{(i)}$
and then rename $\mathbf{x}_{\text{new}}^{(i)}$ as $\mathbf{x}^{(i)}$
for convenience,
\begin{eqnarray}
\langle \Psi^{(i)} |V_{\text{g}}^{\text{pbc},(i)}| \Psi'^{(i)} \rangle
&=& 
\sum_{ \Delta \mathbf{b}^{(i)}} \sum_{\mathbf{b}'^{(i)}}
\int_{-Lb_1'^{(i)}}^{L-Lb_1'^{(i)}} \cdots \int_{-Lb_N'^{(i)}}^{L-Lb_N'^{(i)}}
\exp \biggl[ -\frac{1}{2} (\mathbf{x}^{(i)}-L\Delta \mathbf{b}^{(i)})^T
A^{(i)} (\mathbf{x}^{(i)}-L\Delta \mathbf{b}^{(i)}) \nonumber \\
&&-\frac{1}{2}(\mathbf{x}^{(i)}-\mathbf{s}^{(i)}-L\Delta \mathbf{b}^{(i)})^{T} 
B^{(i)} (\mathbf{x}^{(i)}-\mathbf{s}^{(i)}-L\Delta \mathbf{b}^{(i)}) \biggr]
\nonumber \\
&&
\left[
\sum_{q^{(i)}}   \exp \left(-
\frac{(x_a^{(i)}-x_b^{(i)}+L{b'_a}^{(i)}-L{b'_b}^{(i)}
-Lq^{(i)})^2}{2r_{0}^2}\right)
\right]
\nonumber \\
&&     
\exp \left[ -\frac{1}{2} (\mathbf{x}^{(i)})^T 
A'^{(i)} (\mathbf{x}^{(i)})
-\frac{1}{2}(\mathbf{x}^{(i)}-\mathbf{s}'^{(i)})^{T} 
B'^{(i)} (\mathbf{x}^{(i)}-\mathbf{s}'^{(i)})\right]
d \mathbf{x}^{(i)}.
\end{eqnarray}
Next, we replace
$q^{(i)}-{b'_a}^{(i)}+{b'_b}^{(i)}$ 
by $q_{\text{new}}^{(i)}$.
Since ${b'_a}^{(i)}$ and ${b'_b}^{(i)}$ are fixed,
both $q_{\text{new}}^{(i)}$ and $q^{(i)}$
run through all integers. 
This implies that we can replace the sum over $q^{(i)}$
by a sum over $q_{\text{new}}^{(i)}$.
We then rewrite
$q_{\text{new}}^{(i)}$ as $q^{(i)}$ for convenience,
\begin{eqnarray}
\langle \Psi^{(i)} |V_{\text{g}}^{\text{pbc},(i)}| \Psi'^{(i)} \rangle
&=& 
\sum_{ \Delta \mathbf{b}^{(i)}} \sum_{\mathbf{b}'^{(i)}}
\int_{-Lb_1'^{(i)}}^{L-Lb_1'^{(i)}} \cdots \int_{-Lb_N'^{(i)}}^{L-Lb_N'^{(i)}}
\exp \biggl[ -\frac{1}{2} (\mathbf{x}^{(i)}-L\Delta \mathbf{b}^{(i)})^T
A^{(i)} (\mathbf{x}^{(i)}-L\Delta \mathbf{b}^{(i)}) \nonumber \\
&&-\frac{1}{2}(\mathbf{x}^{(i)}-\mathbf{s}^{(i)}-L\Delta \mathbf{b}^{(i)})^{T} 
B^{(i)} (\mathbf{x}^{(i)}-\mathbf{s}^{(i)}-L\Delta \mathbf{b}^{(i)}) \biggr]
\left[
\sum_{q^{(i)}}   \exp \left(-
\frac{(x_a^{(i)}-x_b^{(i)}
-Lq^{(i)})^2}{2r_{0}^2}\right)
\right]
\nonumber \\
&&     
\exp \left[ -\frac{1}{2} (\mathbf{x}^{(i)})^T 
A'^{(i)} (\mathbf{x}^{(i)})
-\frac{1}{2}(\mathbf{x}^{(i)}-\mathbf{s}'^{(i)})^{T} 
B'^{(i)} (\mathbf{x}^{(i)}-\mathbf{s}'^{(i)})\right]
d \mathbf{x}^{(i)}.
\end{eqnarray}
Lastly, changing the sum over $\mathbf{b}'^{(i)}$ to an integral
and following steps similar to those discussed
in
Sec.~\ref{app_overlap}, we find
\begin{eqnarray}
\langle \Psi^{(i)} |V_{\text{g}}^{\text{pbc},(i)}| \Psi'^{(i)} \rangle
= \nonumber \\
\sum_{\mathbf{b}^{(i)}}
\sum_{q^{(i)}}
\left( \frac{c^{(i)}}{c^{(i)}+2 \rho} \right)^{1/2}
\exp \left\{
-\frac{c^{(i)} \rho}{c^{(i)}+2 \rho}
\left[ \left((C^{(i)})^{-1} \mathbf{v}^{(i)}\right)_{a}-
\left((C^{(i)})^{-1} \mathbf{v}^{(i)}\right)_{b}-Lq^{(i)}\right]^2
\right\} \langle \Psi ^{(i)}
 |\Psi'^{(i)} \rangle,
\end{eqnarray}
where
$(c^{(i)})^{-1}=
[(C^{(i)})^{-1}]_{aa}+[(C^{(i)})^{-1}]_{bb}-
[(C^{(i)})^{-1}]_{ab}-[(C^{(i)})^{-1}]_{ba}$
and
$\rho=1/(2r_{0}^2)$.

\twocolumngrid

\end{document}